# Robust Audio Watermarking Against the D/A and A/D Conversions

Shijun Xiang[1,2],    Jiwu Huang[1,2]

xiangshijun@gmail.com,    isshjw@mail.sysu.edu.cn

1. School of Information Science and Technology, Sun Yat-Sen University, Guangzhou 510275, P. R. China.

2. Guangdong Province Key Laboratory of Information Security, Guangzhou, 510275. P. R. China

**Abstract.** Audio watermarking has played an important role in multimedia security. In many applications using audio watermarking, D/A and A/D conversions (denoted by DA/AD in this paper) are often involved. In previous works, however, the robustness issue of audio watermarking against the DA/AD conversions has not drawn sufficient attention yet. In our extensive investigation, it has been found that the degradation of a watermarked audio signal caused by the DA/AD conversions manifests itself mainly in terms of wave magnitude distortion and linear temporal scaling, making the watermark extraction failed. Accordingly, a DWT-based audio watermarking algorithm robust against the DA/AD conversions is proposed in this paper. To resist the magnitude distortion, the relative energy relationships among different groups of the DWT coefficients in the low-frequency sub-band are utilized in watermark embedding by adaptively controlling the embedding strength. Furthermore, the resynchronization is designed to cope with the linear temporal scaling. The time-frequency localization characteristics of DWT are exploited to save the computational load in the resynchronization. Consequently, the proposed audio watermarking algorithm is robust against the DA/AD conversions, other common audio processing manipulations, and the attacks in StirMark Benchmark for Audio, which has been verified by experiments.

**Indexing keywords:** Audio Watermarking, D/A and A/D Conversions, Resynchronization, Wave Magnitude Distortion, Linear Temporal Scaling, Wavelet Transform.

# 1. Introduction

Audio watermarking [1, 2] plays an important role in ownership protection. According to IFPI (International Federation of the Phonographic Industry) [3], audio watermarking, at a certain data payload sometimes also referred to as data embedding capacity more than 20 bps (Bits Per-Second), should be able to resist the most common signal processing manipulations and attacks, such as temporal scaling, noise corruption, MP3 compression, re-sampling, re-quantization, DA/AD conversions under the constraint of imperceptibility (*SNR*, Signal-to-Noise Ratio, should be higher than 20 dB).

Audio watermarking may be implemented in the time domain or frequency domain. The most typical algorithms in the time domain include LSB-based schemes [4], echo hiding [5], and etc. The algorithms in the frequency domain embed watermark signal by modifying transform coefficients to improve the robustness. Transformations commonly used include DFT (Discrete Fourier Transform) [6, 7], DWT (Discrete Wavelet Transform) [8] and DCT (Discrete Cosine Transform) [9]. Since the watermarked audio is susceptible to shifting and cropping (such as editing, signal interruption in wireless transmission and data packet loss in IP network), the synchronization codes are introduced into audio watermarking [8-11]. In [8, 11], the time-frequency localization characteristics of DWT are investigated and utilized to save the computational cost in the resynchronization and the watermark robustness is improved through data embedding carried out in the low-frequency sub-band of DWT coefficients. In [10], the relationships among different audio sample segments have been used to embed the watermark. This is an effective strategy to resist modification of signal amplitude. Because data is embedded in the time domain, the watermark thus generated cannot achieve good robustness performance.

Today's audio watermark embedding and detection strategies often depend on digital channels such as CD, MP3 and IP network transmission. From the sense of application, however, the robustness of audio watermarking against the DA/AD is an important issue [12]. For instances, in many applications [13-16], watermarks are required to survive in analog environments, during which the DA/AD may be involved. For instance, secret data is proposed to be transmitted via analog telephone channel in [13], and hidden watermark signal may be used to identify pirated music through speaker and PC soundcard, for broadcast music [14-15] or live concert [16] monitoring.

While some audio watermarking algorithms [10, 13-18] are claimed to be robust to the DA/AD conversions, the *BERs* (Bit Error Rate) in these references are either not reported or rather high.

Furthermore, there are no technical descriptions on how to resist the DA/AD conversions. Specifically, none of them have reported how to resist the changes caused by the DA/AD conversions in details. According to the references in [6, 19-20], the serious degradation of audio signal caused by the DA/AD conversions include modification of amplitude and phase, which still puzzles the watermark extraction. Therefore, the DA/AD conversions are considered a challenging issue for audio watermarking [19]. It is also noted that there are no test functions in StirMark Benchmark for Audio [22, 28] that have been designed for evaluating the robustness of audio watermarking to the DA/AD conversions. In summary, to the authors' best knowledge, among all of the literature on audio watermarking, only [19] has discussed the effects caused by the DA/AD conversions on audio, but has not provided the theoretical deduction and experimental investigation for the degradations caused by the DA/AD conversions on audio watermarking, such as waveform distortion and phase modification.

Via extensive investigations, in this paper we analyze and conclude that the main degradation of audio watermark caused by the DA/AD conversions is wave magnitude distortion and linear temporal scaling. Based on these distortions, we propose the following strategies accordingly. Firstly, to resist the wave magnitude distortion of a watermarked audio signal, we adopt the embedding strategy using the relative energy relationships among different groups of the DWT in the low-frequency sub-band and adaptively controlling the embedding strength. This is robust to resist common signal processing manipulations, especially wave magnitude distortion. Secondly, the temporal resynchronization via using synchronization codes is designed to resist the linear temporal scaling. The watermark is resynchronized by computing the scaling factor between two synchronization codes. By exploiting the time-frequency localization characteristics of DWT, the computational load for searching synchronization codes is dramatically saved. Experimental results have demonstrated that our proposed algorithm is robust to the DA/AD conversions, other common audio processing manipulations, and the attacks in StirMark Benchmark for Audio.

The rest of this paper is organized as follows. Section 2 analyzes the effects of the DA/AD conversions on audio watermarking. In Section 3 we introduce the framework and ideas of the proposed algorithm. Section 4 discusses data extraction. Section 5 is the performance analysis for the proposed algorithm. In Section 6, we present the experimental results to demonstrate the performance of the proposed algorithm robust to the DA/AD conversions and other kinds of attacks in StirMark Benchmark for Audio. Finally, the conclusions are drawn in Section 7.

## 2. Effects of the DA/AD conversions on Audio Watermarking

In this section, the effect of the DA/AD conversions on audio watermarking is investigated. In real applications, there are different kinds of transmission channels [21]. In this paper, we are focusing on the effect of the DA/AD conversions without the consideration of analog channels.

### 2.1. Test Scenario

In order to investigate the effects caused by the DA/AD conversions on audio signals, we have designed and used the following test model, as shown in Fig.1. The analog channel may be considered clear here by using a cable line.

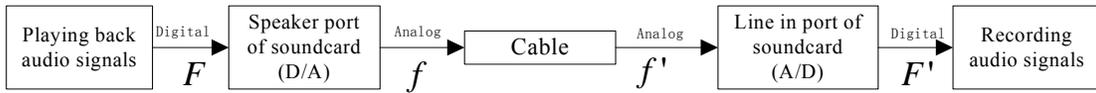

**Fig. 1.** Test model for the DA/AD conversions

**Table 1.** Four different to-be-tested audio files

| File name | Lengths (Sec.) | Properties |
|---|---|---|
| march.wav | 56 | composition of both low and high-frequency |
| drum.wav | 56 | mainly composed of low frequency |
| flute.wav | 56 | well-proportioned frequency distribution |
| dialog.wav | 56 | daily dialog |

Based on the test model in Fig.1, we test many different audio files based on four different 16-bit signed mono audio files in the WAVE format with different frequency properties, denoted by *march.wav*, *drum.wav*, *flute.wav* and *dialog*.wav, sampled at 8, 11.025, 16, 22.05, 32, 44.1, 48, 96 and 128kHz for the purpose of testing, respectively, as listed in Table 1. The *dialog.wav* is about a daily dialog while others three are music generated by the respective music instruments, such as drum, flute, etc. Audio files are played with the tool *Window Media Player 9.0*, after the DA conversion caused by a soundcard the transformed analog signal is output via the *Speaker port* to the *Line in* port of the same or other soundcard, and recorded by the professional tool *CoolEdit 2.1*.

**2.2. Effects of the DA/AD conversions on Audio**

During the DA/AD conversions, digital audio signal will suffer from the following factors [19].

 i. Noise produced by soundcards during DA conversion.

 ii. Modification of audio signal energy and noise energy.

 iii. Noise in analog channel.

 iv. Noise produced by soundcard during the AD conversion including quantization distortion.

Obviously, the digital audio will be distorted after the DA/AD conversions. However, the above knowledge for these distortions is not expressly helpful to design robust watermarking against the DA/AD conversions, we need to further investigate and model them. The detail is described below.

**2.2.1. Linear Temporal Scaling**

Based on the test model shown in Fig.1, numerous different soundcards are employed to test audio files with different sampling frequencies described in Table 1. The experimental results regarding time-scaling during the DA/AD conversions as shown in Table 2. *Sound Blaster Live5.1* is a consumer grade sound board, *ICON StudioPro7.1* is professional one, *Realtek AC'97 audio for VIA (R) Audio controller*, *Audio PCI* and *SoundMAX Digital Audio* are common PC sound blaster. From Table 2, it is noted that during the DA/AD conversions, the time scaling is existing. The detailed observations are described as follows.

i. The scaling factor is different for different soundcards. In other words, during the DA/AD conversions, different performance of soundcards will cause different time-scale distortion.

ii. The scaling factor is also related to the sampling rate of audio files. For the same soundcard, different sampling rate of audio file will have a different scaling distortion during the DA/AD conversions, referred to Table 2. As for other sampling rates (e.g., 22.5 kHz), the conclusion is similar.

iii. The temporal scaling is similar to using two different soundcards are used at a time (One is used to perform the DA conversion while the other for the AD processing).

iv. According to our observations, the scaling factor is in [0, 0.005] for different soundcards and different sampling rates of different kinds of audio files.

**Table 2.** The number of samples changed with the sampling rate of 8 and 44.1 kHz, respectively

| Soundcard / Time | | Blaster Live5.1 | Realtek AC'97 | Audio 2000 PCI | Studio Pro 7.1 | SoundMAX Digital Audio |
|---|---|---|---|---|---|---|
| **8 kHz** | 10s | Reduce: 1 | Increase: 5 | Increase: | Reduce: 70 | Increase: 1 |
| | 20s | Reduce: 2 | Increase: 10 | Increase: | Reduce: 140 | Increase: 2 |
| | 30s | Reduce: 3 | Increase: 15 | Increase: | Reduce: 210 | Increase: 3 |
| | 40s | Reduce: 4 | Increase: 20 | Increase: | Reduce: 280 | Increase: 4 |
| | 50s | Reduce: 5 | Increase: 25 | Increase: | Reduce: 350 | Increase: 5 |
| **44.1 kHz** | 10s | Reduce: 6 | Increase: 4 | Increase: 0 | Increase: 0 | Increase: 2 |
| | 20s | Reduce: 12 | Increase: 8 | Increase: 0 | Increase: 0 | Increase: 4 |
| | 30s | Reduce: 18 | Increase: 12 | Increase: 0 | Increase: 0 | Increase: 6 |
| | 40s | Reduce: 24 | Increase: 16 | Increase: 0 | Increase: 0 | Increase: 8 |
| | 50s | Reduce: 30 | Increase: 20 | Increase: 0 | Increase: 0 | Increase: 10 |

### 2.2.2. Wave Magnitude Distortion

Besides linear temporal scaling discussed above, another kind of degradation on the digital audio caused by the DA/AD conversions is wave magnitude distortion, which is mainly represented as *the modification of signal energy* and *additional noise corruption*. This has been verified by our numerous experiments. It is observed that the audio amplitudes are modified during the DA/AD conversions, and the amount of the modification relies on the volume played back, and the performance of soundcard. Fig.2 and Fig.3 have the same scaling in both horizontal and vertical axis in displaying waves of the original audio and the corresponding recorded audio by the *Blaster Live5.1* soundcard. Compared with the original audio, the amplitude of the recorded audio is obviously reduced. Modification of the amplitude varies with different soundcards.

We use *SNR* to measure the distortion of the recorded audios versus the original audios here. The *SNR* between *F* and *F″* can be expressed as

$$SNR = -10 \log_{10} \left( \frac{\sum_{i=1}^{N} [f(i) - f'(i)]^2}{\sum_{i=1}^{N} [f(i)]^2} \right) \quad (1)$$

$$f''(i) = f'(i) \times \frac{\sum_{i=0}^{N-1} |f(i)|}{\sum_{i=0}^{N-1} |f'(i)|} \quad (2)$$

where $F = \{f(0), f(1),... f(N-1)\}$, $F' = \{f'(0), f'(1),... f'(N-1)\}$ and $F'' = \{f''(0), f''(1),... f''(N-1)\}$ are the original audio, the recorded audio and the normalized audio by Equation (2), respectively. $f(i)$, $f'(i)$ and $f''(i)$ are amplitude of the $i^{th}$ sample of $F$, $F'$ and $F''$, respectively. Note that in the case of linear temporal scaling, $F''$ should be resynchronized (refer to Section 4.1) to generate the resynchronized audio, $F_1''$. Finally, *SNR*s between the original audios and the recorded audios after normalization and resynchronization are calculated.

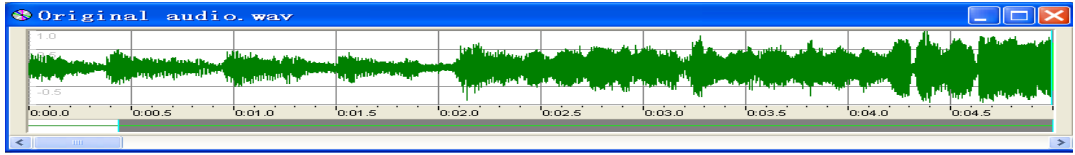

**Fig. 2.** The original audio

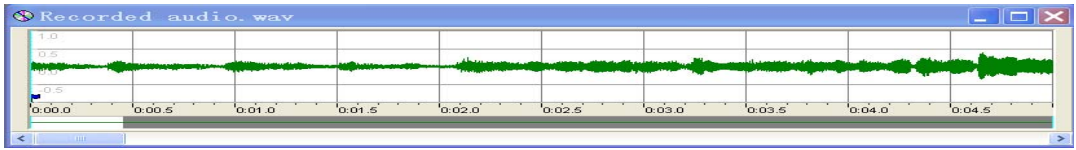

**Fig. 3.** The recorded audio

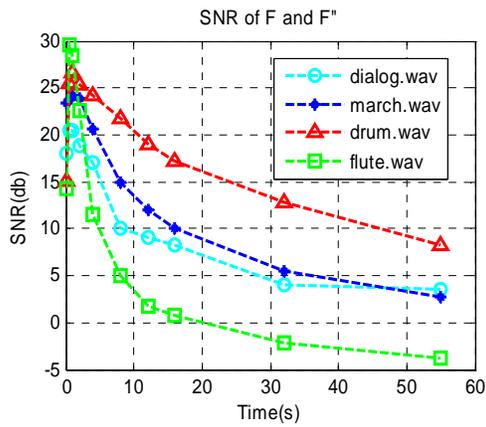 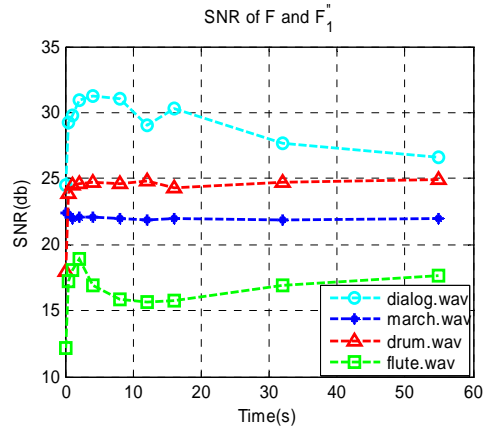

**Fig. 4.** *SNR* curves before resynchronization  F**ig. 5.** *SNR* curves after resynchronization

For experimental description, we choose the soundcard *Sound Blaster Live5.1* and audio files in Table 1 sampled at 44.1 kHz, to test the wave magnitude distortion caused by the DA/AD conversions based on the test model in Fig.1. *SNR*s of $F$ versus $F''$ and $F_1''$ are calculated shown in Figs.4 and 5, respectively.

Fig.4 shows that *SNR*s of some different audio signals $F''$ with respect to $F$ decrease quickly,

demonstrating that linear temporal scaling changed the locations of some samples and thus resulted in serious distortion. And Fig.5 shows the *SNR*s of some different audio signals $F_1''$ with respect to *F*, indicating that the resynchronization effectively resists linear temporal scaling.

### 2.3. Effects of the DA/AD Conversions on Audio Watermarking

In Section 2.2, we discussed the distortion of audio signals caused by the DA/AD conversions, including linear temporal scaling and wave magnitude distortion. From signal processing point of view, watermarks are weak signals embedded in strong background, such as digital audio and image. Therefore, any changes of the cover-signal will directly influence the survival of the watermarks. Hence, it is expected that audio watermarking will suffer from the DA/AD conversions in terms of the following two aspects.

i. Linear temporal scaling, meaning that the shift of audio samples in the time domain will directly influence the detection of the embedded watermark.

ii. Audio signal magnitude distortion, meaning that the effect of signal energy modification followed by an additive noise will affect watermark extraction.

Mathematically speaking, the effect of the DA/AD conversions on Audio Watermarking can be formulated as follows,

$$f'(i) = \lambda \cdot f(\frac{i}{\alpha}) + \eta \qquad (3)$$

where $\alpha$ and $\lambda$ denote linear temporal scaling factor and amplitude scaling factor, $\eta$ is additive noise on the sample $f(i)$, and $f'(i)$ is the version of $f(i)$ after the DA/AD Conversions. If $\frac{i}{\alpha}$ is not an integer, $f(\frac{i}{\alpha})$ is interpolated with the nearest samples in the shifted audio signal. Via extensive experiments, a suggesting value of parameters is given: $\alpha \subset [0, 0.005]$, $\lambda \subset [0.5, 2.0]$.

### 3. Proposed Embedding Algorithm

After investigating the effect of linear temporal scaling and wave magnitude distortion caused by the DA/AD conversions, a new watermarking procedure is proposed to combat the degradation due to the DA/AD conversions. Referred to Equation (3), to resist wave magnitude distortion formulated as the effect of the parameters $\lambda$ and $\eta$, the relative energy relationships among different groups of the DWT coefficients in the low-frequency sub-band are used in the embedding stage by adaptively controlling the embedding strength. Furthermore, the resynchronization is designed to cope with linear temporal scaling in the detection by referring to the effect of the parameter $\alpha$. The proposed watermarking strategy is addressed below.

### 3.1. Embedding Framework

The main idea of the proposed embedding algorithm is to split a long audio into many segments, and then embed one synchronization code and a portion of to-be-embedded information bits into the DWT coefficients in the low-frequency sub-band of each segment. The embedding model is shown in Fig.6.

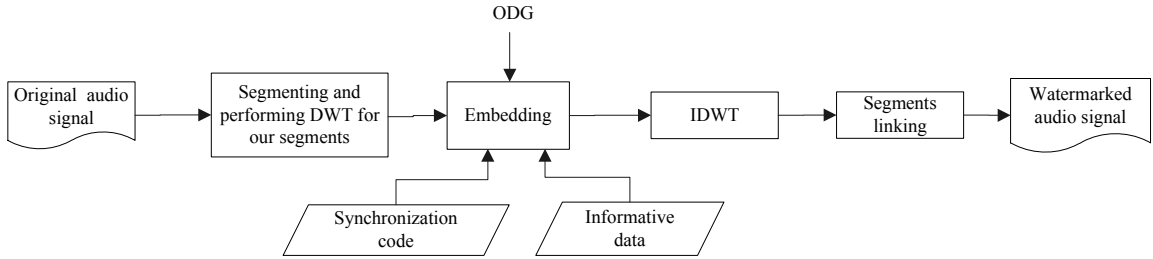

**Fig. 6.** Watermark embedding framework

In the algorithm, the watermark bits are embedded in such a way that the embedding strength is adaptively controlled to achieve the largest value under the imperceptibility constraint (the ODG value is in [0, -2]). The detail is described as follows. Suppose that $S_1$ is the ODG of the watermarked audio versus the original audio, $S_0$, a predefined value. If $S_1 < S_0$, the embedding strength will be automatically decreased until $S_1 \geq S_0$. The ODG is computed in the DWT domain instead of in the time domain. This efficiently reduces the computational load by avoiding the inverse discrete wavelet transform (IDWT) process. After the watermark embedded, the IDWT is performed to reconstruct the watermarked audio.

Usually, the SNR is used to measure the distortion due to the watermark. Since the SNR is not an acceptable measure for audio quality without the consideration of the human auditory system, here the ODG value of PEAQ model is used to control the watermarking distortion instead of common SNR measure referred to Section 6.1, the distortion caused by the watermark is considered acceptable when the ODG value is in [0, -2].

### 3.2. Embedding Strategy

As mentioned above and will be further discussed in the rest of this paper, the proposed embedding algorithm is conducted in DWT domain because of its superiority. To hide data robust to modification of audio amplitude, the watermarks are embedded in DWT domain by using the relative relationships among different groups of the DWT coefficients, $\{c(i)\}$. It is noted that utilizing the relationships among different audio sample sections to embed data is proposed in [10]. However, what proposed in this paper is different from [10]. First, instead of in the time domain, we embed watermark signal in the low-frequency sub-band of DWT in order to achieve better robustness performance. Furthermore, in the DWT domain, the time-frequency localization characteristic of DWT can be exploited to save the computational load during searching for synchronization codes [8, 11]. Denote three consecutive DWT coefficient groups by Group_1, Group _2 and Group _3. Each group includes $L$ coefficients, as shown in Fig.7.

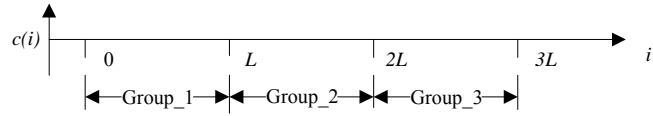

**Fig. 7.** Three consecutive coefficient groups in DWT low-frequency subband

Generally, $L$ is chosen based on the embedding bit rate, *SNR* of the watermarked audio and the embedding strength. The energy of those three coefficient groups, denoted by $E_1$, $E_2$ and $E_3$, respectively, is defined as

$$E_1 = \sum_{i=0}^{L-1} |c(i)|, \quad E_2 = \sum_{i=L}^{2L-1} |c(i)|, \quad E_3 = \sum_{i=2L}^{3L-1} |c(i)| \tag{4}$$

where $c(i)$ is the $i^{th}$ DWT coefficient in the low-frequency subband. So their energy differences may be obtained according to the following equation.

$$\begin{cases} A = E_{max} - E_{med} \\ B = E_{med} - E_{min} \end{cases} \tag{5}$$

where $E_{max} = \max\{E_1, E_2, E_3\}$, $E_{med} = med\{E_1, E_2, E_3\}$, $E_{min} = \min\{E_1, E_2, E_3\}$, and *max*, *med* and *min* are up operations picking the maximum, medium and minimum of $E_1$, $E_2$ and $E_3$. $A$ and $B$ stand for the energy differences. Equation (6) is exploited to express the embedding strength.

$$S = (d \cdot \sum_{i=0}^{3L-1} |c(i)|)/3 \qquad (6)$$

The parameter $d$ in Equation (6) is the embedding strength factor, which plays a role in deciding the embedding strength $S$. To resist wave magnitude distortion during the DA/AD conversions, the value of $d$ should be as large as possible under the imperceptibility constraint (measured by *ODG*). Usually, during embedding watermarks, $d$ is assigned with a predefined value at first, and then modified automatically until *SNR* of the watermarked audio falls into the predefined range. *SNR* reduces as $d$ increasing.

After calculating the energy differences among three DWT coefficient groups by Equation (5), one watermark bit $w(i)$ may be embedded through the relationship between $A$ and $B$, shown as Equation (7).

$$\begin{cases} A - B \geq S & if \ w(i) = 1 \\ B - A \geq S & if \ w(i) = 0 \end{cases} \qquad (7)$$

Note that if the embedded bit $w(i)$ is '1' and $A - B \geq S$ or the embedded bit $w(i)$ is '0' and $B - A \geq S$, then there is no operation. Otherwise, the values of different groups of the DWT coefficients used to compute $E_{max}$, $E_{med}$ and $E_{min}$ will be adjusted until satisfying $A - B \geq S$ or $B - A \geq S$ in terms of the embedded bit, referred to Equation (4). The rules applied to modify $E_{max}$, $E_{med}$ and $E_{min}$ are formulated as Equations (8), (9), (10) and (11).

If the embedded bit $w(i)$ is '1' and $A - B < S$, we apply Equation (8) by modifying the values of different groups of the DWT coefficients to satisfy $A - B \geq S$:

$$c'(i) = \begin{cases} c(i) \times (1 + \dfrac{|\beta|}{E_{max} + 2E_{med} + E_{min}}), & if \ c(i) \ is \ coefs \ used \ to \ generate \ E_{max} \ and \ E_{min} \\ c(i) \times (1 - \dfrac{|\beta|}{E_{max} + 2E_{med} + E_{min}}), & if \ c(i) \ is \ coefs \ used \ to \ generate \ E_{med} \end{cases} \qquad (8)$$

where $|\beta| = |A - B - S| = S - A + B$. Modifying the DWT coefficients to embedding '1' with Equation (8) may cause the case that $E_{med} < E_{min}$ due to the increase of $E_{min}$ and the decrease of $E_{med}$. Obviously, this situation will influence the watermark extraction. In order to remain the relation of $E_{max} \geq E_{med} \geq E_{min}$ after embedding '1', we derive Equation (9) to control the embedding strength.

$$S \leq \dfrac{2E_{med}}{E_{med} + E_{min}}(E_{max} - E_{min}) \qquad (9)$$

Similarly, if $w(i)$ is '0' and $B - A < S$, we apply Equation (10) by modifying the values of different groups of the DWT coefficients to satisfy $B - A \geq S$:

$$c'(i) = \begin{cases} c(i) \times (1 - \dfrac{|\beta|}{E_{\max} + 2E_{med} + E_{\min}}), & \text{if } c(i) \text{ is coefs used to generate } E_{\max} \text{ and } E_{\min} \\ c(i) \times (1 + \dfrac{|\beta|}{E_{\max} + 2E_{med} + E_{\min}}), & \text{if } c(i) \text{ is coefs used to generate } E_{med} \end{cases} \quad (10)$$

where $|\beta| = |B - A - S| = S + A - B$. Also, using Equation (10) to modify the DWT coefficients to embed bit '0' may lead to $E_{med} > E_{\max}$ due to the increase of $E_{med}$ and the decrease of $E_{\max}$. This will influence the watermark extraction. Accordingly, we deduct Equation (11) to elude the case by controlling the embedding strength.

$$S \leq \frac{2E_{med}}{E_{\max} + E_{med}}(E_{\max} - E_{\min}) \quad (11)$$

For the proof of Equations (9) and (11), please refer to APPENDIX I and II.

It is worth noting that our proposed watermarking embedding strategy is largely different from that in [10]. Compared with the method in [10], our proposed strategy efficiently remains the relation of the maximal, middle and minimal unchanged, that is $E_{max} > E_{med} > E_{min}$ before the embedding and after the embedding $E'_{max} > E'_{med} > E'_{min}$. In addition, the computation cost is reduced, i.e. the computation load is $O(3 \times L)$ for our proposed strategy and $O(3 \times L \times M)$ in [10], respectively. $M$, often greater than 1, is the repeated times in the embedding.

### 3.3. Watermarking and Synchronization Code

The algorithm exploits a PN (Pseudo-random Noise) sequence as a synchronization code. The synchronization code is used to locate the position of hidden informative bits, thus resisting the cropping and shifting attacks [8]. In this paper, the synchronization code is introduced to resist linear temporal scaling caused by the DA/AD conversions.

Suppose $\{Syn(i)\}$ is an original synchronization code and $\{Seq(i)\}$ is an unknown sequence both having the same length. If the number of different bits between $\{Syn(i)\}$ and $\{Seq(i)\}$, when compared bit-by-bit, are less than or equal to a predefined threshold, $T$, the sequence $\{Seq(i)\}$ will be determined as the synchronization code. The analysis of error probability in searching synchronization codes is given in

Section 5.2.

To make the watermark robust to linear temporal scaling during the DA/AD conversions, we embed the watermark with synchronization codes. Before embedding, the synchronization codes {*Syn*(*i*)} and watermark {*Wmk*(*i*)} should be arranged into a binary data sequence, as shown in Fig. 8.

According to the length of {*Syn*(*i*)} and {*Wmk*(*i*)}, the original audio is split into proper segments and then perform DWT on each segment. As shown in Fig.8 and Fig.9, the sequences {*Syn*(*i*)} and {*Wmk*(*i*)} are embedded into the DWT low-frequency sub-band of Segment_1 and Segment_2, respectively. We embed one information bit in three DWT coefficient groups see Fig. 7), including 3×*L* coefficients. Let $N_1$ and $N_2$ denote the lengths of samples in Segment_1 and Segment_2 in the time domain. So

$$N_1 = 3L \times 2^K \times (The\ length\ of\ synchronization\ code) \qquad (12)$$

$$N_2 = 3L \times 2^K \times (The\ length\ of\ the\ watermark) \qquad (13)$$

where *K* is the levels of DWT decomposition, and *L* is the number of the DWT coefficients in a group.

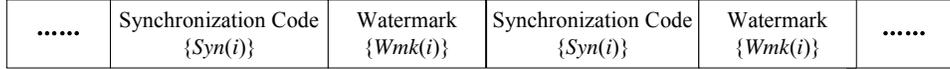

**Fig. 8.** Data structure of hidden bit stream

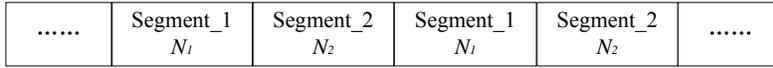

**Fig. 9.** Segmenting of the host audio signal

## 4. Proposed Extraction Algorithm

Linear temporal scaling will modify the positions of audio samples and lead to serous distortion (as discussed in Section 2.2). To resist linear temporal scaling, synchronization codes and the strategy of resynchronization are applied in the extracting. During searching for synchronization codes, the time-frequency localization characteristics of DWT are exploited to save the computational load. The model of data extraction is shown in Fig.10.

### 4.1. Resynchronization in the Time Domain

After the DA/AD conversions, we need to locate the embedded watermark bits by searching for synchronization code in to-be-tested audio, $F'$, and then compute the number of the samples between two synchronization codes, denoted as $N_2'$. Note that the number of the samples $N_2$ in the original watermarked audio between two synchronization codes is a predefined value (refer to Section 3.3). Since linear temporal scaling occurs repeatedly during the DA/AD conversions (refer to Section 2.2.1), a scaling factor is defined to perform the resynchronization, described as follows.

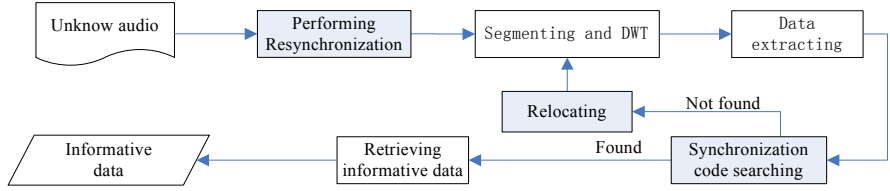

**Fig. 10.** Block diagram of data extraction

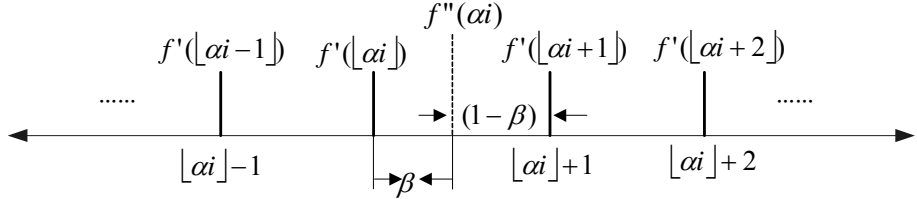

**Fig. 11.** The sketch map of resynchronization operation in Equation (15).

Suppose that $F = \{f(i) \mid i = 0, 1, \cdots, N_2 - 1\}$ and $F' = \{f'(0), f'(1), \ldots f'(N_2'-1)\}$ are the original watermarked audio and the audio after the DA/AD conversions between two synchronization codes, respectively. We define the scaling factor as

$$\alpha = N_2'/N_2 \qquad (14)$$

In searching synchronization codes, the computational load has been dramatically reduced by using the time-frequency localization characteristics of DWT. Thus it resolves the contending requirements between robustness of hidden data and efficiency of synchronization codes searching.

The resynchronization is a procedure based on the interpolation processing, which is designed to descale the linear temporal scaling. We have tested a few kinds of interpolation formulas (such as Lagrange, Newton, etc.), and the simulation results are similar in terms of resisting the linear temporal scaling. Referred to Fig. 11, here, we list the most efficient Lagrange linear interpolation as follows,.

$$f"(i) = \begin{cases} f'(0), & if \quad i = 0, \\ (1-\beta) \cdot f'(\lfloor \alpha \cdot i \rfloor) + \beta \cdot f'(\lfloor \alpha \cdot i \rfloor + 1), & if \quad 0 < i < N_2 - 1, \\ f'(N_2'-1), & if \quad i = N_2 - 1 \end{cases} \quad (15)$$

where $F" = \{f"(0), f"(1), \ldots f"(N_2 - 1)\}$ is the version of $F'$ after synchronization manipulation, $f"(i)$ and $f'(j)$ the $i^{th}$ and $j^{th}$ sample of $F"$ and $F'$, $0 \leq i \leq N_2 - 1$, $0 \leq j \leq N_2'-1$, $\lfloor \ \rfloor$ the floor function. $\beta = \alpha \cdot i - \lfloor \alpha \cdot i \rfloor$.

### 4.2. Data Extraction

After resynchronization, we perform the same DWT on the resynchronized audio segment $F"$ as in the embedding to obtain the low-frequency sub-band coefficients $\{c"(i)\}$. As in Equation (4), we compute $E_1"$, $E_2"$ and $E_3"$. Where, $E_1" = \sum_{i=0}^{L-1} |c"(i)|$, $E_2" = \sum_{i=L}^{2L-1} |c"(i)|$, $E_3" = \sum_{i=2L}^{3L-1} |c"(i)|$, then which are ordered to obtain $E"_{max}$, $E"_{med}$ and $E"_{min}$. Similar to Equation (5), we defined

$$\begin{cases} A" = E"_{max} - E"_{med} = \max\{E_1", E_2", E_3"\} - med\{E_1", E_2", E_3"\} \\ B" = E"_{med} - E"_{min} = med\{E_1", E_2", E_3"\} - \min\{E_1", E_2", E_3"\} \end{cases} \quad (16)$$

Comparing $A"$ and $B"$, we get the retrieved bit by using the following rule.

$$w"(i) = \begin{cases} 1 & if \quad A" - B" \geq 0 \\ 0 & if \quad A" - B" < 0 \end{cases} \quad (17)$$

The process is repeated to retrieve the whole binary data stream.

In the extracting, the synchronization sequence $\{Seq(i)\}$ and the number of samples predefined between two synchronization codes $N_2$ are beforehand known, and the original DWT coefficients are not required in the extracting. Thus, watermarking detection is blind.

### 5. Performance Analysis

In this section, we evaluate the performance of the proposed algorithm in terms of data embedding capacity or payload, error probability of synchronization codes and watermarks, and resisting amplitude modification attack. The *BER* is defined as

$$BER = \frac{Number\ of\ error\ bits}{Number\ of\ total\ bits} \times 100\% \quad (18)$$

Because we use the orthogonal wavelet in our algorithm and the embedding rules in Section 3.2 do not change the DWT coefficients in the high-frequency subbands of $F$, we rewrite Equation (1) as

$$SNR = -10\log_{10}\left(\frac{\|F-F'\|_2^2}{\|F\|_2^2}\right) = -10\log_{10}\left(\frac{\|C-C'\|_2^2}{\|F\|_2^2}\right) \quad (19)$$

where $C = \{c_i\}$ and $C' = \{c'_i\}$ are the DWT coefficients in low frequency sub-band of $F$ and $F'$ after a $K$-level DWT.

### 5.1. Data Embedding Capacity

The data embedding capacity refers to the number of bits that are embedded into the audio signal within a unit of time, measured in the unit of *bps* and denoted by $B$. Suppose that the sampling rate of audio is $R$ (Hz). The data embedding capacity $B$ of the proposed algorithm can be expressed as

$$B = R/(3 \cdot L \cdot 2^K) \quad (bps) \quad (20)$$

where $K$ and $L$ denotes wavelet decomposition level and the number of the DWT coefficients in a group.

### 5.2. Error Analysis on Synchronization Code Searching

There are two types of errors in searching synchronization codes, false positive error and false negative error. A false positive error occurs when a synchronization code is supposed to be detected in the location where no synchronization code is embedded, while a false negative error occurs when an existing synchronization code is missed. Once a false positive error occurs, the bits after the locations of the false synchronization code will be regarded as the watermark bits. When a false negative error takes place, some watermark bits will be lost. The false positive error probability of the synchronization code $P_1$ can be calculated as follows:

$$P_1 = \frac{1}{2^{N_1}} \cdot \sum_{k=0}^{T} C_{N_1}^k \quad (21)$$

where $N_1$ is the length of a synchronization code, and $T$ is the threshold introduced in Section 3.3.

Generally, we use the following formulation to evaluate the false negative error probability $P_2$ of the synchronization code according to the bit error probability, denoted as $P_d$, in the detector.

$$P_2 = \sum_{k=T+1}^{N_1} C_{N_1}^k \cdot (P_d)^k \cdot (1-P_d)^{N_1-k} \qquad (22)$$

In our works, how to resynchronize the watermark is an important issue after the DA/AD conversions since the watermark is tracked by using synchronization codes. Therefore, we need to consider the robustness of synchronization codes to the linear temporal scaling caused by the DA/AD conversions. In [11], the authors proposed a strategy by increasing the redundancy of the synchronization bits and watermark bits locally to improve the robustness of the watermark and successfully resist pitch-invariant TSM (Time-Scale Modification) attacks of 4%. Specifically, an 8-bit synchronization sequence with the local redundancy rate 3, 10101011, is defined as 111000111000111000111111. Actually, the redundancy is a simple style of ECC (Error Correcting Codes) [30]. In most of cases during DA/AD conversions, the amount of the linear temporal scaling is very small, especially at 44.1 kHz or higher only several even no samples shifted (referred to Section 2.2.1). In other words, it is a feasible way to improve the robustness of the synchronization code through using the local redundancy. In our extensive experiments, the synchronization code with the length of 31 bits is robust to the linear temporal scaling caused by the DA/AD conversions.

**5.3. Error Analysis on Watermark Extraction**

It is noted that the introduction of synchronization codes in the algorithm may make the difference between the bit error probability of the watermark in the detector $P_d$ and in the channel $P_w$, illustrated in Fig.12.

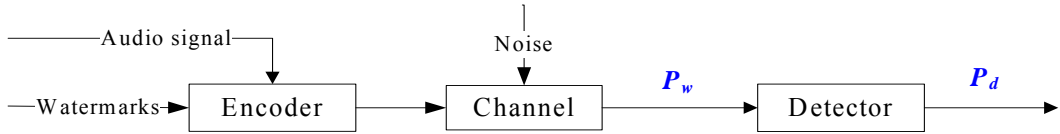

**Fig. 12**. The sketch map of the watermark bit error probability in the channel and detector

Supposed that $x$, the number of synchronization codes, are embedded and the number of the false positive synchronization codes and false negative synchronization codes detected is $y$ and $z$, respectively. So the error probability $P_w$ may be expressed as follows. The false positive error probability $P_1$ can be expressed as $y/(x-z+y)$ here.

$$P_w = \frac{(x-z) \cdot N_2 \cdot P_{sw} + y \cdot N_2 \cdot P_{aw}}{(x+y-z) \cdot N_2} = (1-P_1) \cdot P_{sw} + P_1 \cdot P_{aw} \qquad (23)$$

In Equation (23), $N_2$ is the length of the watermark bits, which follow a corresponding synchronization code, $P_{sw}$ and $P_{aw}$ denote the error probability of the watermarks in case of false negative and positive synchronization code occurring. From the view of point in probability theory, the value of $P_{sw}$ and $P_{aw}$ is approximately $P_d$ and 50%. Accordingly, we have the following formulation.

$$P_w = (1-P_1) \cdot P_{sw} + P_1 \cdot P_{aw} \approx (1-P_1) \cdot P_d + P_1 \cdot 50\% \qquad (24)$$

From Equation (24), it is noted that the bit error probability of the watermark in the channel is different from that in the detector after introducing synchronization code, and the difference mainly relies on the number of the false positive synchronization code. The occurring of the false negative synchronization code will lead to the loss of some hidden information bits, the effect of which on the error probability of the watermark may be ignored. When the value of $y$ go to ZERO, $P_1$ goes to ZERO, thus $P_w$ goes to $P_d$.

**5.4. Amplitude Modification Attack**

Some audio signal processing or hostile attacks may modify the audio amplitudes, such as the wave magnitude distortion caused by the DA/AD conversion, which may be considered as amplitude scaling followed by the additional noise. Refer to Equation (4), the sums of different groups of the DWT coefficients after amplitude modification attack may be formulated as $E'_{max} = \varphi \cdot E_{max} + \Delta_1$, $E'_{med} = \varphi \cdot E_{med} + \Delta_2$, $E'_{min} = \varphi \cdot E_{min} + \Delta_3$. So, using Equation (5) the difference between $A'$ and $B'$ after the amplitude modification attack is computed as

$$\begin{cases} A' - B' = E'_{max} - 2E'_{med} + E'_{min} = \varphi \cdot (E_{max} - 2E_{med} + E_{min}) + \Delta_1 - 2\Delta_2 + \Delta_3 \\ B' - A' = 2E'_{med} - E'_{max} - E'_{min} = \varphi \cdot (2E_{med} - E_{max} - E_{min}) + 2\Delta_2 - \Delta_1 - \Delta_3 \end{cases} \qquad (25)$$

where $\varphi$ is amplitude scaling factor, $\Delta_1$, $\Delta_2$ and $\Delta_3$ the maximum, medium and minimum of the additional noise in three consecutive DWT coefficient groups, respectively. By using Equations (5), (6), (7) and (17), we may conclude the conditions to correctly extract the watermark bit $w(i)$, defined as follows,

$$\begin{cases} \Delta_1 - 2\Delta_2 + \Delta_3 \geq -\varphi \cdot (E_{max} - 2E_{med} + E_{min}) \geq -\varphi \cdot \sigma & if \quad w(i) = 1 \\ \Delta_1 - 2\Delta_2 + \Delta_3 \leq \varphi \cdot (2E_{med} - E_{max} - E_{min}) < \varphi \cdot \sigma & if \quad w(i) = 0 \end{cases} \qquad (26)$$

where $S \leq E_{max} - 2E_{med} + E_{min} \leq \sigma$ if $w(i) = 1$, and $S \leq 2E_{med} - E_{max} - E_{min} \leq \sigma$ if $w(i) = 0$. $S$ is the

embedding strength, a positive number.

For amplitude scaling attack, $\Delta_1 = \Delta_2 = \Delta_3 = 0$ and $\varphi \cdot \sigma \geq 0$, indicating that $w(i)$ can be always detected correctly by Equation (26), meaning that the watermark is robust to amplitude scaling attack.

## 6. Experimental Results

In the following experiments, each synchronization code is composed of a 31 bit m-sequence and the corresponding watermark is a binary sequence with 32 bits. Six decomposition levels of db2 wavelet are applied. The length of the DWT coefficient group $L$ (refer to Fig.7) is 8. With Equation (20), we can estimate the data embedding capacity as 28.71 bps for 44.1 kHz of sampling rate. It needs an audio segment about 2.2 seconds to embed a synchronization code and a watermark. Totally 25 synchronization codes and watermarks are embedded in audio of 56 seconds, the length of the watermarks 800 bits.

### 6.1. Quality Evaluation of Watermarked Audio

The *SNRs* between the original audio and the watermarked audios are controlled as 20+ dB, satisfying the IFPI requirement, and the watermarks are not imperceptible during listening test. Since the SNR values are definitely NOT a good imperceptibility measure, here we also provide with more relevant imperceptibility results through the software EAQUAL 0.1.3 alpha [23-24], which considers HAS (human auditory system) models. EAQUAL is coded by the ITU-R recommendation BS.1387 [25] to stand for watermarked audio quality. Table 3 shows the value of MOVs (Model Output Variables), and the corresponding *SNRs* of audio files are 23.67 (*march.wav*), 21.67 (*drum.wav*), 29.97 (*flute.wav*) and 15.63 (*dialog.wav*).

In Table 3, the NMR (Noise-To-Mask-Ratiovalue) of the files *drum.wav* and *dialog.wav* is respectively 15.94 and 4.23 more than 1.5 dB while their corresponding ODG (Objective Difference Grade) value is -3.91 and -3.77, very close to -4. The audio quality ratings are measured with the five point scale defined in ITU-R BS.1116 and thus the SDG and ODG have a range of [-4;0] where -4 stands for very annoying difference and 0 stands for imperceptible difference between reference and test signal. In addition, if the NMR of any frequency band is higher 1.5dB the frame is assumed to be disturbed. In this case, the *march.wav* and *flute.wav* is considered inaudible because their ODGs are close to ZERO and NMRs are less than 1.5 dB. And, the *drum.wav* and *dialog.wav* are assumed to be disturbed by the watermarks embedded, but the SNR value of *drum.wav* is an acceptable value 21.67 dB according to IFPI [3], indicating the test standard based on EAQUAL and SNR is largely different. Therefore, adaptive watermark embedding via MOVs instead of SNR is applied in this paper, referred to Fig.6. About the detail descriptions of MOVs, please refer to [23-26].

**6.2. Robustness Performance**

Table 4 shows the robustness of the watermarks embedded in audios with sampling rate of 44.1 kHz against the DA/AD conversions by *Sound Blaster Live5.1* soundcard. Clearly, the introduction of synchronization technique largely reduces *BER*, average from 16.75% to 0.4375%, and further reduced to 0.0625% by using resynchronization manipulation.

Obviously, the proposed algorithm efficiently improves the robustness of the watermark against the DA/AD conversions. It means that the embedding strategy based on relative energy relationships among different groups of the DWT coefficients has resisted wave magnitude distortion caused by the DA/AD conversions and the resynchronization operation based on the interpolation processing has descaled the linear temporal scaling. Accordingly, we believe that the linear temporal scaling caused by the DA/AD conversions may be modeled as an interpolation processing, equivalent to the playback speed modification presented in [29], and the resynchronization operation is the corresponding inverse processing operation.

In the extraction, the number of the false positive synchronization codes and false negative synchronization codes detected is zero, thus according to Equations (23) or (24). So we may use $P_d$ to evaluate the false negative error probability $P_2$ by using Equation (22). To choose the threshold *T*, both the false positive error probability and the false negative error probability should be considered. We choose T to be 5 in our work based on Table 5.

Table 6 shows that our algorithms are very robust to common signal processing manipulations such as mp3 compression, amplitude scaling, re-sampling and re-quantization, low-pass filtering (LPF), and etc. It is owing to that in proposed watermarking scheme the watermark is embedded in the low-frequency component of DWT domain.

Table 7 shows the performance of our algorithm and several recent audio watermarking strategies in [8, 10, 17, 18] against the DA/AD conversions, and some common audio processing manipulations, like Gaussian noise corruption and MP3 compression. These algorithms are implemented and simulated by using the test scenario of Fig.1 and the audio files described in Table 1. Compared with them, our algorithm not only can effectively resist Gaussian noise corruption and MP3 compression, but also achieves higher robustness performance against the DA/AD conversions. It shows that the proposed watermarking method is exactly pertinent to the distortion caused by the DA/AD conversion:

i. The temporal scaling in DA/AD conversions is minor and can be represented as an interpolation processing operation. The minor scaling will shift the sample position not so much, that means the synchronization code can effectively to locate the position of watermark embedding. Due to the scaling is an interpolation processing, making the interpolation-based resynchronization can descale its effect.

ii. The embedding strategy based on the relative relation is immune to the energy change in the DA/AD conversion;

iii. The corruption due to additive noise in the DA/AD processing can be combated by embedding the watermark in the low-frequency sub-band of DWT domain.

Consequently, the watermark robust to the DA/AD conversions is achieved. In order to further evaluate the performance of our proposed algorithm, we report the experimental results regarding common audio signal processing, and Stirmark Benchmark for Audio with the different kinds of audio files described in Section 2. The test results are similar for different clips. Here, we take the mono signed audio file *march.wav* with sampling rate of 44.1 kHz as a example to present the performance of proposed watermarking strategy, as tabulated in Tables 7-8. The audio editing and attacking tools adopted in experiment are CoolEdit Pro v2.1, Goldwave v5.10 and Stirmark for Audio v0.2.

Table 8 is the experimental results with "Stirmark for Audio" v0.2, indicating that the watermark is robust to most attacks. However, it is noted that our proposed algorithm is sensitive to others kinds of Stirmark attacks, such as VoiceRemove, AddFFTNoise, FFT_HLPass, RC_HighPass, CopySample, FFT_Test and FFT_stat1 attack. There are mainly the following reasons.

i. Firstly, the watermark is removed by destroying the content of the watermarked audio. For instance, after VoiceRemove, AddFFTNoise processing, listening testing shows that the content of the audio is almost ruined completely.

ii. Secondly, the watermark is removed by deleting the low-frequency band of the cover-signal as the result of the attack FFT_HLPass or RC_HighPass. In this paper, since the watermark is embedded into the low-frequency sub-band of the DWT coefficients, the hidden information is removed completely as long as the low-frequency component of watermarked signals is removed.

iii. Thirdly, the watermark is removed because the energy relationships among sample points are modified. Since the watermark is embedded with the relative relationship strategy, as a result of the FFT_Test and FFT_stat1 attack, the energy relationships among different groups of the DWT coefficients are modified due to swapping samples in FFT domain, resulting in the watermark extraction failed.

iv. The proposed algorithm is also robust to random removal even cutting one every 10 samples, but sensitive to the CopySample attack, which also changed the energy relationship among samples. When the *BER* is 20+%, we think that the watermark extraction is failed.

Robustness, imperceptibility and capacity of a watermark affect each other. In proposed watermarking algorithm, the robustness relies on the level of DWT decomposition. Their relation is the stronger robustness, the lower data embedding capacity and the severer the distortion due to the watermark.

**Table 3.** Quality Evaluation of the Watermarked Audio Using EQUAL

| Audio Files<br>Movs | *march.wav* | *drum.wav* | *flute.wav* | *dialog.wav* |
|---|---|---|---|---|
| **BandwidthRef** | 20704.7882 | 10111.9922 | 17751.6744 | 10667.8519 |
| **BandwidthTest** | 20704.7882 | 10111.9922 | 17751.6118 | 10667.8402 |
| **Total NMR** | -16.2629 | 15.9422 | -22.5610 | 4.2279 |
| **WinModDiff1** | 4.8290 | 116.0379 | 3.7197 | 35.2723 |
| **ADB** | 0.7421 | 2.4325 | 0.6324 | 2.1682 |
| **EHS** | 0.1150 | 1.2770 | 0.1443 | 0.7374 |
| **AvgModDiff1** | 4.2554 | 117.3014 | 3.1457 | 13.9112 |
| **AvgModDiff2** | 12.4491 | 1489.2157 | 6.9712 | 33.7151 |
| **RmsNoiseLoud** | 0.1106 | 2.0097 | 0.2760 | 1.6434 |
| **MFPD** | 1.0000 | 1.0000 | 1.0000 | 1.0000 |
| **RDF** | 0.1086 | 0.8893 | 0.0124 | 0.5498 |
| **DIX** | 2.24 | -4.09 | 2.69 | -2.95 |
| **ODG** | -0.19 | -3.91 | -0.05 | -3.77 |

**Table 4.** The bits error of the watermarks in audios in the DA/AD conversions

| Audio Files | | *march.wav* | *drum.wav* | *flute.wav* | *dialog.wav* | Average |
|---|---|---|---|---|---|---|
| No Synchronization Code | Error Bits | 137/800 | 174/800 | 191/800 | 34/800 | 134/800 |
| | *BER*(%) | 17.12 | 21.75 | 23.88 | 4.25 | 16.75 |
| No Resynchronization | Error Bits | 0 | 4/800 | 7/800 | 2/800 | 3.25/800 |

|  | BER(%) | 0 | 0.5 | 0.875 | 0.25 | 0.4375 |
|---|---|---|---|---|---|---|
| Our algorithm | Error Bits | 0 | 0 | 2/800 | 0 | 0.5/800 |
|  | BER(%) | 0 | 0 | 0.25 | 0 | 0.0625 |

Here, 3.25 is the average value of error bits related to different test clips, (0+4+7+2)/4=3.25.

**Table 5.** The relationships among $T$, $P_1$ and $P_2$

| T | | $T = 5$ | $T = 6$ | $T = 7$ | $T = 8$ |
|---|---|---|---|---|---|
| No Resynchronization | $P_1$ | $9.61 \times 10^{-5}$ | $4.39 \times 10^{-4}$ | $5.30 \times 10^{-3}$ | $1.70 \times 10^{-3}$ |
|  | $P_2$ | $4.70 \times 10^{-9}$ | $7.36 \times 10^{-11}$ | $1.09 \times 10^{-14}$ | $9.68 \times 10^{-13}$ |
| Our algorithm | $P_1$ | $9.61 \times 10^{-5}$ | $4.39 \times 10^{-4}$ | $5.30 \times 10^{-3}$ | $1.70 \times 10^{-3}$ |
|  | $P_2$ | $4.33 \times 10^{-14}$ | $9.67 \times 10^{-17}$ | $2.90 \times 10^{-22}$ | $1.81 \times 10^{-19}$ |

**Table 6.** Robustness Performance to Common Attacks

| Attack Type | BER(%) | Attack Type | BER(%) |
|---|---|---|---|
| Unattacked | 0 | Gaussain (8 dB) | 0 |
| MP3 (32 kbps) | 0 | MP3 (128 kbps) | 0 |
| Requantization (8 bit) | 0 | Resample (8 kHz) | 0 |
| LPF (LowPassFreq = 9000 Hz) | 0 | Amplitude scaling (10%~150%) | 0 |

**Table 7.** The performance of different algorithms

| Algorithm | Data embedding capacity (bps) | Gaussian noise BER(%) | MP3 Compression BER(%) | Resisting the DA/AD conversions BER(%) |
|---|---|---|---|---|
| Ref. [8] | About 172 | 0 (8 dB) | 0 (32 kbps) | Failed to extracting |
| Ref.[10] | About 49 | Not mentioned | About 2.92 (80 kbps) | About 2 |
| Ref. [17] | About 8.53 | 2.73 (36 dB) | About 2.99 (64 kbps) | About 1.3 |
| Ref. [18] | About 25 | Not mentioned | About 1.42 (64 kbps) | About 3.57 |
| Our | About 28.71 | 0 (8 dB) | 0 (32 kbps) | About 0.0625 |

**Table 8.** Robustness Performance to Other kinds of Attacks in Stirmark for Audio v0.2

| Attack Type | BER(%) | Parameters |
|---|---|---|
| AddBrumm_100 | 0 | AddBrummFreq = 55, AddBrummfrom = 100 |
| AddBrumm _1100 | 15.79 | AddBrummto = 10100, AddBrummstep = 1000 |
| AddNoise_100 | 0 | Noisefrom = 100 |
| AddNoise _500 | 0.5 | Noiseto = 1000 |
| AddNoise _900 | 5.875 | Noisestep = 200 |
| Compressor | 0 | ThresholdDB = -6.123, CompressValue = 2.1 |
| AddSinus | 0 | AddSinusFreq = 900, AddSinusAmp = 1300 |
| AddDynNoise | 0 | Dynnoise = 20 |
| Amplify | 0 | Amplify = 50 |
| Exchange | 0 | |
| ExtraStereo_30 | 0 | ExtraStereofrom = 30 |
| ExtraStereo _50 | 0 | ExtraStereoto = 70 |
| ExtraStereo _70 | 0 | ExtraStereostep = 20 |
| Normalize | 0 | |
| ZeroLength | 0 | ZeroLength = 10 |
| ZeroCross | 0 | ZeroCross = 1000 |
| Invert | 0 | |
| Nothing | 0 | |
| Original | 0 | |
| Stat1 | 0 | |
| Stat1 | 0 | |
| RC _LowPass | 0 | LowPassFreq = 9000 |
| Smooth2 | 0 | |
| Smooth | 0 | |
| FFT_Invert | 0 | FFTSIZE = 16384 |
| FFT_RealReverse | 0 | FFTSIZE = 16384 |
| ZeroRemove | 0 | |
| Echo | 0 | Period = 10 |
| Echo | 13.04 | Period = 50 |
| FlippSample | 0 | Period = 10, FlippDist = 6, FlippCount = 2 |
| FlippSample | 19.5 | Period = 1000, FlippDist = 600, FlippCount= 200 |
| CutSample | 0 | Remove = 10, RemoveNumber = 1 |
| CopySample | 19.97 | Period = 10, FlippDist = 6, FlippCount = 1 |
| FFT_Test | Failed | FFTSIZE = 16384 |
| AddFFTNoise | Failed | FFTSIZE = 16384, FFTNoise = 30000 |
| FFT _HLPass | Failed | FFTSIZE = 16384, HighPassFreq =200,LowPassFreq = 9000 |
| FFT _Stat1 | Failed | |
| RC_HighPass | Failed | HighPassFreq = 200 |
| VoiceRemove | Failed | |

# 7. Conclusions

We are proposing in this paper an audio watermarking scheme aiming at solving the DA/AD conversions without the consideration of analog channels. The main content is simply concluded as follows:

i. Based on many experiments, we analyze the effects of the DA/AD conversions on the audio, and conclude the main degradations on the audio watermark are linear temporal scaling and wave magnitude distortion. The amount of the linear temporal scaling relies on the tested soundcards and sampling rate of the tested audio. Compared with TSM algorithms, the temporal scaling caused by the DA/AD conversions is linear and usually minor. And, the linear temporal scaling caused by the DA/AD conversions may be considered as a resample interpolation processing manipulation verified by experiments. The wave magnitude distortion may be modeled as amplitude scaling followed by the additional noise corruption.

ii. By analyzing the distortion caused by the DA/AD conversions, we propose a pertinent watermarking strategy. To resist wave magnitude distortion, we adopt the embedding strategy of the relative energy relationships among different groups of the DWT coefficients in the low-frequency sub-band by adaptively controlling the embedding strength. Furthermore, the interpolation-based resynchronization via synchronization codes is designed to resist the linear temporal scaling. By exploiting the time-frequency localization characteristics of DWT, the computational cost for resynchronization is dramatically saved, and the robustness of the watermark is improved by embedding in the low-frequency sub-band.

iii. By introducing the synchronization code, resynchronization strategy and DWT technique, we present a DWT-based blind audio watermark algorithm robust to the DA/AD conversions by referring to the effect of the DA/AD on audio signals. We also evaluate the performance of the proposed algorithm in terms of data embedding capacity, error probability of synchronization code and watermark extraction, and resisting amplitude modification attack. The experimental results show the watermark is very robust against the DA/AD conversions, most of common audio processing operations or the attacks

In this paper, we focus on the investigation on the DA/AD conversion. The effect of different analog transmission channels [21] with more DA/AD conversion devices will be a consideration of future works.

## Acknowledgments

This work is supported by NSFC (60325208, 60403045), NSF of Guangdong (04205407), New Jersey Commission of Science and Technology via New Jersey Center of Wireless Networking and Internet Security (NJWINS).

## References

[1] M. Arnold, "Audio watermarking: features, applications and algorithms," *Proc. of IEEE Int. Conf. on Multimedia & Expo*, New York, USA, vol. 2, pp. 1013-1016, 2000.

[2] M. D. Swanson, B. Zhu, A. H. Tewfik, "Current state of the art, challenges and future directions for audio watermarking," *Proc. of IEEE Int. Conf. on Multimedia Computing and Systems*, vol. 1, pp.19-24, 1999.

[3] S. Katzenbeisser, F. A. P. Petitcolas, ed. *Information Hiding Techniques for Steganography and Digital Watermarking*. Artech House, Inc., 2000.

[4] M. A. Gerzon, P. G. Graven, "A high-rate buried-data channel for audio CD," *Journal of the Audio Engineering Society*, vol. 43, pp. 3-22, 1995.

[5] D. Gruhl, A. Lu, W. Bender, "Echo hiding," *Proc. of the 1$^{st}$ Information Hiding Workshop*. LNCS, vol. 1174, Berlin: Germany Springer-Verlag, pp. 295–315, 1996.

[6] W. Bender, D. Gruhl, N. Morimoto, "Techniques for data hiding," *IBM Systems Journal*, vol. 35, pp. 313–336, 1996.

[7] Sang-Kwang Lee, Yo-Sung Ho, "Digital audio watermarking in the cepstrum domain," *IEEE Trans. on Consumer Electronics*, vol. 46, pp. 744-750, 2000.

[8] S. Q. Wu, J. W. Huang, D. R. Huang, Y. Q. Shi, "Efficiently self-synchronized audio watermarking for assured audio data transmission," *IEEE Trans. on Broadcasting*, vol. 51, no. 1, pp. 69-76, 2005.

[9] J. W. Huang, Y. Wang, Y. Q. Shi, "A blind audio watermarking algorithm with self-synchronization," *Proc. of IEEE Int. Sym. on Circuits and Systems*, vol. 3, pp. 627-630, 2002.

[10] W.-N Lie, L.-C. Chang, "Robust and high-quality time-domain audio watermarking subject to

psychoacoustic masking," *Proc. of IEEE Int. Sym. on Circuits and Systems*, vol. 2, pp. 45-48, 2002.

[11] H. O. Kim, B. K. Lee, N. Y. Lee, "Wavelet-based audio watermarking techniques: robustness and fast synchronization". http://amath.kaist.ac.kr/research/paper/01-11.pdf

[12] C. I. Podilchuk and E. J. Delp, "Digital watermarking: algorithms and applications," *IEEE Signal Processing Magazine*, vol. 18, pp. 33-46, July 2001.

[13] S. Chen, H. Leung, "Concurrent data transmission on analog telephone channel by data hiding technique," *Proc. of IEEE Int. Sym. on Consumer Electronics*, pp.295-298, 2004.

[14] J. Haitsma, M. van der Veen, T. Kalker, F. Bruekers, "Audio watermarking for monitoring and copy protection," *Proc. of ACM Multimedia Workshops*, pp. 119-122, 2000.

[15] T. Nakamura, R. Tachibana, and S. Kobayashi, "Automatic music monitoring and boundary detection for broadcast using audio watermarking," *Proc. SPIE*, vol. 4675, pp. 170-180, 2002.

[16] R. Tachibana, "Audio watermarking for live performance," *Proc. SPIE*, vol. 5020, pp. 32-43, 2003.

[17] J. Seok, J. Hong, and J. Kim, "A novel audio watermarking algorithm for copyright protection of digital audio," *ETRI Journal*, vol. 24, no.3, pp. 181-189, 2002.

[18] S. Shin, O. Kim, J. Kim, J. Choil, "A robust audio watermarking algorithm using pitch scaling," *Proc. of IEEE Workshop on Digital Signal Processing*, vol. 2, pp. 701–704, 2002.

[19] M. Steinebach, A. Lang, J. Dittmann, C. Neubauer, "Audio watermarking quality evaluation: robustness to DA/AD processes," *Proc. of Int. Conf. on Information Technology: Coding and Computing*, pp. 100-103, 2002.

[20] R. Popa, "An Analysis of Steganographic Techniques," *Ph. D Thesis*, pp. 26-27, 1998.

[21] S. J. Xiang, J. W. Huang. "Analysis of D/A and A/D conversions in quantization-based audio watermarking," *International Journal of Network Security*, Vol. 3, pp. 230-238, 2006.

[22] Online: Available at: [https://amsl-smb.cs.uni-magdeburg.de/smfa//allgemeines.php] .

[23] Online: Available at: [http://www.mp3-tech.org/programmer/sources/eaqual.tgz] .

[24] Online: Available at: [http://www.mp3-tech.org/programmer/misc.html].

[25] International Telecommunication Union, "Method for Objective Measurements of Perceived Audio Quality (PEAQ)," *ITU-R BS 1387*, 1998.

[26] M. Arnold, "Subjective and objective quality evaluation of watermarked audio tracks," *Web Delivering of Music*, pp. 161-167, 2002.

[27] W. Li, X. Y. Xue and P. Z. Lu, "Robust audio watermarking based on rhythm region Detection," *IEE Electronics Letters*, vol. 41, no. 4, pp. 75-76, 2005.

[28] M. Steinebach, F.A.P. Petitcolas, F. Raynal, J. Dittmann, C. Fontaine, S. Seibel, N. Fates, L.C. Ferri, "StirMark benchmark: audio watermarking attacks," *Proc. of Int. Conf. on Information Technology: Coding and Computing*, pp. 49-54, 2001.

[29] B. Sylvain, V. D. V. Michiel and L. Aweke, "Informed detection of audio watermark for resolving Playback speed modifications," *Proc. the Multimedia and Security Workshop*, pp. 117-123, 2004.

[30] L.H. Charles Lee, ed. Error-Control Block Codes for Communications Engineers. Artech House, Inc., 2000.

# APPENDIX I

When the embedding bit is '1' and $A - B < S$, we modify the values of different groups of the DWT coefficients to satisfy $A - B \geq S$ according to Equation (8). Let $\beta = A - B - S$, so

$$E'_{max} = E_{max} \times (1 + \frac{|\beta|}{E_{max} + E_{min} + 2E_{med}}), \ E'_{min} = E_{min} \times (1 + \frac{|\beta|}{E_{max} + E_{min} + 2E_{med}}), \ E'_{med} = E_{med} \times (1 - \frac{|\beta|}{E_{max} + E_{min} + 2E_{med}}).$$

Where $E'_{max}, E'_{med}, E'_{min}$ is the corresponding version of $E_{max}, E_{med}, E_{min}$. Obviously, after embedding bit '1', $E'_{max}, E'_{min}$ increased while $E'_{med}$ decreased on the basis of $E_{max}, E_{med}, E_{min}$. This may generate $E'_{min} > E'_{med}$. According Equations (5), (6) and (7), we have $|\beta| = S + B - A = S + 2E_{med} - E_{max} - E_{min}$. In order to avoid the case after embedding bit '1', we have the following deduction.

$$\begin{aligned}
& E'_{med} \geq E'_{min} \\
\Leftrightarrow \ & E_{med} \times (1 - \frac{|\beta|}{E_{max} + E_{min} + 2E_{med}}) \geq E_{min} \times (1 + \frac{|\beta|}{E_{max} + E_{min} + 2E_{med}}) \\
\Leftrightarrow \ & E_{med} \times (E_{max} + E_{min} + 2E_{med} - |\beta|) \geq E_{min} \times (E_{max} + E_{min} + 2E_{med} + |\beta|) \\
\Leftrightarrow \ & E_{med} \times (2E_{max} + 2E_{min} - S) \geq E_{min} \times (4E_{med} + S) \\
\Leftrightarrow \ & S \times (E_{med} + E_{min}) \leq (2E_{max}E_{med} - 2E_{med}E_{min}) \\
\Leftrightarrow \ & S \leq \frac{2E_{med}}{E_{med} + E_{min}}(E_{max} - E_{min})
\end{aligned}$$

The proof of Equation (9) is finished.

# APPENDIX II

When the embedding bit is '0' and $B - A < S$, we modify the values of different groups of the DWT coefficients to satisfy $B - A \geq S$ according to Equation (10). Let $\beta = B - A - S$, so

$$E'_{max} = E_{max} \times (1 - \frac{|\beta|}{E_{max} + E_{min} + 2E_{med}}), \quad E'_{min} = E_{min} \times (1 - \frac{|\beta|}{E_{max} + E_{min} + 2E_{med}}), \quad E'_{med} = E_{med} \times (1 + \frac{|\beta|}{E_{max} + E_{min} + 2E_{med}}).$$

Where $E'_{max}, E'_{med}, E'_{min}$ is the corresponding version of $E_{max}, E_{med}, E_{min}$. Obviously, after embedding bit '0', $E'_{max}, E'_{min}$ decreased while $E'_{med}$ increased on the basis of $E_{max}, E_{med}, E_{min}$. This may lead to $E'_{med} > E'_{max}$. According Equations (5), (6) and (7), we have $|\beta| = S + A - B = S + E_{max} - 2E_{med} + E_{min}$. In order to avoid the case occurring after embedding bit '0', we have the following deduction.

$$\begin{aligned} &E'_{max} \geq E'_{med} \\ \Leftrightarrow\ &E_{max} \times (1 - \frac{|\beta|}{E_{max} + E_{min} + 2E_{med}}) \geq E_{med} \times (1 + \frac{|\beta|}{E_{max} + E_{min} + 2E_{med}}) \\ \Leftrightarrow\ &E_{max} \times (E_{max} + E_{min} + 2E_{med} - |\beta|) \geq E_{med} \times (E_{max} + E_{min} + 2E_{med} + |\beta|) \\ \Leftrightarrow\ &E_{max} \times (4E_{max} - S) \geq E_{med} \times (2E_{min} + 2E_{max} + S) \\ \Leftrightarrow\ &S \times (E_{med} + E_{max}) \leq (2E_{max}E_{med} - 2E_{med}E_{min}) \\ \Leftrightarrow\ &S \leq \frac{2E_{med}}{E_{med} + E_{max}}(E_{max} - E_{min}) \end{aligned}$$

The proof of Equation (11) is finished.